\documentclass[aps,prd,reprint,superscriptaddress,altaffilletter]{revtex4-1}
\usepackage{graphicx}
\usepackage{hyperref}
\usepackage{color}

\begin{document}

\preprint{5.11}

\title{Search for nucleon decays with EXO-200}
\collaboration{EXO-200 Collaboration}\noaffiliation


\author{J.B.~Albert}\affiliation{Physics Department and CEEM, Indiana University, Bloomington, Indiana 47405, USA}
\author{G.~Anton}\affiliation{Erlangen Centre for Astroparticle Physics (ECAP), Friedrich-Alexander-University Erlangen-N\"urnberg, Erlangen 91058, Germany}
\author{I.~Badhrees}\altaffiliation{Permanent address: King Abdulaziz City for Science and Technology, Riyadh, Saudi Arabia}\affiliation{Physics Department, Carleton University, Ottawa, Ontario K1S 5B6, Canada}
\author{P.S.~Barbeau}\affiliation{Department of Physics, Duke University, and Triangle Universities Nuclear Laboratory (TUNL), Durham, North Carolina 27708, USA}
\author{R.~Bayerlein}\affiliation{Erlangen Centre for Astroparticle Physics (ECAP), Friedrich-Alexander-University Erlangen-N\"urnberg, Erlangen 91058, Germany}
\author{D.~Beck}\affiliation{Physics Department, University of Illinois, Urbana-Champaign, Illinois 61801, USA}
\author{V.~Belov}\affiliation{Institute for Theoretical and Experimental Physics, Moscow, Russia}
\author{M.~Breidenbach}\affiliation{SLAC National Accelerator Laboratory, Menlo Park, California 94025, USA}
\author{T.~Brunner}\affiliation{Physics Department, McGill University, Montr\'{e}al, Qu\'{e}bec H3A 2T8,  Canada}\affiliation{TRIUMF, Vancouver, British Columbia V6T 2A3, Canada}
\author{G.F.~Cao}\affiliation{Institute of High Energy Physics, Beijing 100049, China}
\author{W.R.~Cen}\affiliation{Institute of High Energy Physics, Beijing 100049, China}
\author{C.~Chambers}\affiliation{Physics Department, Colorado State University, Fort Collins, Colorado 80523, USA}
\author{B.~Cleveland}\altaffiliation{Also at SNOLAB, Sudbury, Ontario, Canada}\affiliation{Department of Physics, Laurentian University, Sudbury, Ontario P3E 2C6, Canada}
\author{M.~Coon}\affiliation{Physics Department, University of Illinois, Urbana-Champaign, Illinois 61801, USA}
\author{A.~Craycraft}\affiliation{Physics Department, Colorado State University, Fort Collins, Colorado 80523, USA}
\author{W.~Cree}\affiliation{Physics Department, Carleton University, Ottawa, Ontario K1S 5B6, Canada}
\author{T.~Daniels}\affiliation{Department of Physics and Physical Oceanography, University of North Carolina at Wilmington, Wilmington, North Carolina 28403, USA}
\author{M.~Danilov}\altaffiliation{Present address: P.N.Lebedev Physical Institute of the Russian Academy of Sciences, Moscow, Russia}\affiliation{Institute for Theoretical and Experimental Physics, Moscow, Russia}
\author{S.J.~Daugherty}\affiliation{Physics Department and CEEM, Indiana University, Bloomington, Indiana 47405, USA}
\author{J.~Daughhetee}\affiliation{Department of Physics, University of South Dakota, Vermillion, South Dakota 57069, USA}
\author{J.~Davis}\affiliation{SLAC National Accelerator Laboratory, Menlo Park, California 94025, USA}
\author{S.~Delaquis}\affiliation{SLAC National Accelerator Laboratory, Menlo Park, California 94025, USA}
\author{A.~Der~Mesrobian-Kabakian}\affiliation{Department of Physics, Laurentian University, Sudbury, Ontario P3E 2C6, Canada}
\author{R.~DeVoe}\affiliation{Physics Department, Stanford University, Stanford, California 94305, USA}
\author{T.~Didberidze}\affiliation{Department of Physics and Astronomy, University of Alabama, Tuscaloosa, Alabama 35487, USA}
\author{J.~Dilling}\affiliation{TRIUMF, Vancouver, British Columbia V6T 2A3, Canada}
\author{A.~Dolgolenko}\affiliation{Institute for Theoretical and Experimental Physics, Moscow, Russia}
\author{M.J.~Dolinski}\affiliation{Department of Physics, Drexel University, Philadelphia, Pennsylvania 19104, USA}
\author{W.~Fairbank Jr.}\affiliation{Physics Department, Colorado State University, Fort Collins, Colorado 80523, USA}
\author{J.~Farine}\affiliation{Department of Physics, Laurentian University, Sudbury, Ontario P3E 2C6, Canada}
\author{S.~Feyzbakhsh}\affiliation{Amherst Center for Fundamental Interactions and Physics Department, University of Massachusetts, Amherst, Massachusetts 01003, USA}
\author{P.~Fierlinger}\affiliation{Technische Universit\"at M\"unchen, Physikdepartment and Excellence Cluster Universe, Garching 80805, Germany}
\author{D.~Fudenberg}\affiliation{Physics Department, Stanford University, Stanford, California 94305, USA}
\author{R.~Gornea}\affiliation{Physics Department, Carleton University, Ottawa, Ontario K1S 5B6, Canada}\affiliation{TRIUMF, Vancouver, British Columbia V6T 2A3, Canada}
\author{K.~Graham}\affiliation{Physics Department, Carleton University, Ottawa, Ontario K1S 5B6, Canada}
\author{G.~Gratta}\affiliation{Physics Department, Stanford University, Stanford, California 94305, USA}
\author{C.~Hall}\affiliation{Physics Department, University of Maryland, College Park, Maryland 20742, USA}
\author{E.V.~Hansen}\affiliation{Department of Physics, Drexel University, Philadelphia, Pennsylvania 19104, USA}
\author{J.~Hoessl}\affiliation{Erlangen Centre for Astroparticle Physics (ECAP), Friedrich-Alexander-University Erlangen-N\"urnberg, Erlangen 91058, Germany}
\author{S.~Homiller}\affiliation{Physics Department, University of Illinois, Urbana-Champaign, Illinois 61801, USA}
\author{P.~Hufschmidt}\affiliation{Erlangen Centre for Astroparticle Physics (ECAP), Friedrich-Alexander-University Erlangen-N\"urnberg, Erlangen 91058, Germany}
\author{M.~Hughes}\affiliation{Department of Physics and Astronomy, University of Alabama, Tuscaloosa, Alabama 35487, USA}
\author{A.~Jamil}\altaffiliation{Present address: Yale University, New Haven, Connecticut, USA}\affiliation{Erlangen Centre for Astroparticle Physics (ECAP), Friedrich-Alexander-University Erlangen-N\"urnberg, Erlangen 91058, Germany}\affiliation{Physics Department, Stanford University, Stanford, California 94305, USA}
\author{M.J.~Jewell}\affiliation{Physics Department, Stanford University, Stanford, California 94305, USA}
\author{A.~Johnson}\affiliation{SLAC National Accelerator Laboratory, Menlo Park, California 94025, USA}
\author{S.~Johnston}\altaffiliation{Present address: Argonne National Laboratory, Argonne, Illinois USA}\affiliation{Amherst Center for Fundamental Interactions and Physics Department, University of Massachusetts, Amherst, Massachusetts 01003, USA}
\author{A.~Karelin}\affiliation{Institute for Theoretical and Experimental Physics, Moscow, Russia}
\author{L.J.~Kaufman}\altaffiliation{Present address: SLAC National Accelerator Laboratory, Menlo Park, California, USA}\affiliation{Physics Department and CEEM, Indiana University, Bloomington, Indiana 47405, USA}
\author{T.~Koffas}\affiliation{Physics Department, Carleton University, Ottawa, Ontario K1S 5B6, Canada}
\author{S.~Kravitz}\altaffiliation{Present address: Lawrence Berkeley National Laboratory, Berkeley, California, USA}\affiliation{Physics Department, Stanford University, Stanford, California 94305, USA}
\author{R.~Kr\"{u}cken}\affiliation{TRIUMF, Vancouver, British Columbia V6T 2A3, Canada}
\author{A.~Kuchenkov}\affiliation{Institute for Theoretical and Experimental Physics, Moscow, Russia}
\author{K.S.~Kumar}\affiliation{Department of Physics and Astronomy, Stony Brook University, SUNY, Stony Brook, New York 11794, USA}
\author{Y.~Lan}\affiliation{TRIUMF, Vancouver, British Columbia V6T 2A3, Canada}
\author{D.S.~Leonard}\affiliation{IBS Center for Underground Physics, Daejeon 34047, Korea}
\author{G.S.~Li}\affiliation{Physics Department, Stanford University, Stanford, California 94305, USA}
\author{S.~Li}\affiliation{Physics Department, University of Illinois, Urbana-Champaign, Illinois 61801, USA}
\author{C.~Licciardi}\altaffiliation{Present address: Laurentian University, Sudbury, Canada}\affiliation{Physics Department, Carleton University, Ottawa, Ontario K1S 5B6, Canada}
\author{Y.H.~Lin}\affiliation{Department of Physics, Drexel University, Philadelphia, Pennsylvania 19104, USA}
\author{R.~MacLellan}\affiliation{Department of Physics, University of South Dakota, Vermillion, South Dakota 57069, USA}
\author{T.~Michel}\affiliation{Erlangen Centre for Astroparticle Physics (ECAP), Friedrich-Alexander-University Erlangen-N\"urnberg, Erlangen 91058, Germany}
\author{B.~Mong}\affiliation{SLAC National Accelerator Laboratory, Menlo Park, California 94025, USA}
\author{D.~Moore}\affiliation{Department of Physics, Yale University, New Haven, Connecticut 06511, USA}
\author{K.~Murray}\affiliation{Physics Department, McGill University, Montr\'{e}al, Qu\'{e}bec H3A 2T8, Canada}
\author{R.~Nelson}\affiliation{Waste Isolation Pilot Plant, Carlsbad, New Mexico 88220, USA}
\author{O.~Njoya}\affiliation{Department of Physics and Astronomy, Stony Brook University, SUNY, Stony Brook, New York 11794, USA}
\author{A.~Odian}\affiliation{SLAC National Accelerator Laboratory, Menlo Park, California 94025, USA}
\author{I.~Ostrovskiy}\affiliation{Department of Physics and Astronomy, University of Alabama, Tuscaloosa, Alabama 35487, USA}
\author{A.~Piepke}\affiliation{Department of Physics and Astronomy, University of Alabama, Tuscaloosa, Alabama 35487, USA}
\author{A.~Pocar}\affiliation{Amherst Center for Fundamental Interactions and Physics Department, University of Massachusetts, Amherst, Massachusetts 01003, USA}
\author{F.~Reti\`{e}re}\affiliation{TRIUMF, Vancouver, British Columbia V6T 2A3, Canada}
\author{A.L.~Robinson}\affiliation{Department of Physics, Laurentian University, Sudbury, Ontario P3E 2C6, Canada}
\author{P.C.~Rowson}\affiliation{SLAC National Accelerator Laboratory, Menlo Park, California 94025, USA}
\author{S.~Schmidt}\affiliation{Erlangen Centre for Astroparticle Physics (ECAP), Friedrich-Alexander-University Erlangen-N\"urnberg, Erlangen 91058, Germany}
\author{A.~Schubert}\altaffiliation{Present address: OneBridge Solutions, Boise, Idaho, USA}\affiliation{Physics Department, Stanford University, Stanford, California 94305, USA}
\author{D.~Sinclair}\affiliation{Physics Department, Carleton University, Ottawa, Ontario K1S 5B6, Canada}\affiliation{TRIUMF, Vancouver, British Columbia V6T 2A3, Canada}
\author{A.K.~Soma}\affiliation{Department of Physics and Astronomy, University of Alabama, Tuscaloosa, Alabama 35487, USA}
\author{V.~Stekhanov}\affiliation{Institute for Theoretical and Experimental Physics, Moscow, Russia}
\author{M.~Tarka}\affiliation{Department of Physics and Astronomy, Stony Brook University, SUNY, Stony Brook, New York 11794, USA}
\author{T.~Tolba}\affiliation{Institute of High Energy Physics, Beijing 100049, China}
\author{R.~Tsang}\altaffiliation{Present address: Pacific Northwest National Laboratory, Richland, Washington, USA}\affiliation{Department of Physics and Astronomy, University of Alabama, Tuscaloosa, Alabama 35487, USA}
\author{P.~Vogel}\affiliation{Kellogg Lab, Caltech, Pasadena, California 91125, USA}
\author{J.-L.~Vuilleumier}\affiliation{LHEP, Albert Einstein Center, University of Bern, Bern, Switzerland}
\author{M.~Wagenpfeil}\affiliation{Erlangen Centre for Astroparticle Physics (ECAP), Friedrich-Alexander-University Erlangen-N\"urnberg, Erlangen 91058, Germany}
\author{A.~Waite}\affiliation{SLAC National Accelerator Laboratory, Menlo Park, California 94025, USA}
\author{T.~Walton}\altaffiliation{Present address: Prism Computational Sciences, Madison, Wisconsin, USA}\affiliation{Physics Department, Colorado State University, Fort Collins, Colorado 80523, USA}
\author{M.~Weber}\affiliation{Physics Department, Stanford University, Stanford, California 94305, USA}
\author{L.J.~Wen}\affiliation{Institute of High Energy Physics, Beijing 100049, China}
\author{U.~Wichoski}\affiliation{Department of Physics, Laurentian University, Sudbury, Ontario P3E 2C6, Canada}
\author{G.~Wrede}\affiliation{Erlangen Centre for Astroparticle Physics (ECAP), Friedrich-Alexander-University Erlangen-N\"urnberg, Erlangen 91058, Germany}
\author{L.~Yang}\affiliation{Physics Department, University of Illinois, Urbana-Champaign, Illinois 61801, USA}
\author{Y.-R.~Yen}\affiliation{Department of Physics, Drexel University, Philadelphia, Pennsylvania 19104, USA}
\author{O.Ya.~Zeldovich}\affiliation{Institute for Theoretical and Experimental Physics, Moscow, Russia}
\author{J.~Zettlemoyer}\affiliation{Physics Department and CEEM, Indiana University, Bloomington, Indiana 47405, USA}
\author{T.~Ziegler}\affiliation{Erlangen Centre for Astroparticle Physics (ECAP), Friedrich-Alexander-University Erlangen-N\"urnberg, Erlangen 91058, Germany}



\date{\today}

\begin{abstract}
A search for instability of nucleons bound in $^{136}$Xe nuclei is reported with 223 kg$\cdot$yr exposure of $^{136}$Xe in the EXO-200 experiment. Lifetime limits of 3.3$\times 10^{23}$ and 1.9$\times 10^{23}$~yrs are established for nucleon decay to $^{133}$Sb and $^{133}$Te, respectively. These are the most stringent to date, exceeding the prior decay limits by a factor of 9 and 7, respectively.
\end{abstract}


\maketitle


\section{Introduction}\label{sec:intro}
Current experimental data are consistent with baryon number (B) and lepton number (L) conservation. However, proton stability is not guaranteed by a fundamental symmetry. If baryon number is not an exact 
symmetry, its violation would have a deep effect on the understanding of the evolution of
the Universe, in particular on the origin of the matter-antimatter asymmetry.

The Standard Model (SM) successfully explains most experimental data at energies below a few hundred GeV, yet it is generally regarded as an effective field theory valid only up to some cut-off scale $\Lambda$. In many extensions of the SM, baryon and lepton numbers are no longer conserved. An example of a process violating only lepton number conservation (by two units) is neutrinoless double-beta decay ($0\nu\beta\beta$), which may occur in several even-even nuclei, but has not been observed yet. Violation of total lepton number by two units could be related to the dimension 5 operator, the so-called Weinberg operator, \(\frac{llHH}{\Lambda_L}\) (where $l$ is the left-handed lepton doublet, $H$ is the Higgs doublet, and $\Lambda_L$ is the cut-off scale associated with lepton number violation). This is the lowest dimension operator that can produce neutrinoless double-beta decay. The EXO-200 experiment has searched for the signatures of this process in \(^{136}\)Xe for the two most commonly considered mechanisms - decays with the emission of two electrons only~\cite{Albert:2014b} and decays with the additional emission of one or two massive bosons, called Majorons~\cite{EXOMajoron:2014}. With the ``natural'' assignments of parameters, the current limits on the $0\nu\beta\beta$ decay rate~\cite{Albert:2014b,Gando:2016,Gerda:2017} test $\Lambda_L$ up to scales of 10$^{14}$ to 10$^{15}$ GeV. Analogously, the dimension 6 operator \(\frac{QQQl}{\Lambda^2_B}\) (where $Q$ is quark doublet and $\Lambda_B$ is the cut-off scale associated with baryon number violation) would cause both B and L violation. The current limits~\cite{SK:2014,SK:2009,IMB:1999} on protons decaying into $\pi^0e^+$ and $K^+\bar{\nu}$ provide limits on $\Lambda_B$ in the range of 10$^{15}$ to 10$^{16}$ GeV, similar to the ones obtained for $\Lambda_L$.

Nevertheless, it is possible that there are symmetries suppressing the simple baryon $\Delta$B = 1 and lepton number $\Delta$L = 2 violating processes despite the fact that the corresponding baryon number violating scale is relatively low. An example of such a symmetry was proposed in~\cite{Babu:2003} where it was shown that the SM Lagrangian, extended to include small neutrino masses, has an anomaly-free Z$_6$ discrete gauge symmetry. This symmetry forbids the well tested $\Delta$B = 1 and $\Delta$B = 2 decay processes but allows the $\Delta$B = 3 triple nucleon decay. This mode of nucleon decay, which may occur through dimension 15 operators, is tested in this work at the $\Lambda \sim$10$^2$ GeV energy scale [9].

\begin{figure}[hbpt]
\centering
\includegraphics[scale = .4]{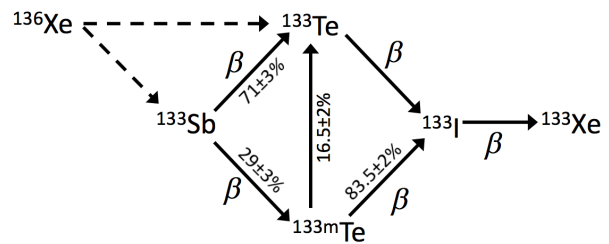}
\caption{Decay chains for daughter nuclei resulting from nucleon decay of $^{136}$Xe. The branching ratios are from \cite{Braun:1984}.}
\label{fig:chain}
\end{figure}

\section{Search strategy}
For nuclei with mass number A $\gg$ 3 four triple nucleon combinations (ppp,
npp, nnp, or nnn) could undergo the $\Delta$B = 3 decay. As a result of this decay, A-3 nucleons will remain, unless additional baryons are emitted by an excited daughter nucleus. The decay chain of the daughter nucleus is a possible signature of the triple nucleon decay. Observing these chains is the strategy already used in~\cite{DAMA:2006} for the case of $^{136}$Xe. An analogous strategy has also been applied to a search for $\Delta$B =1 and $\Delta$B = 2 decays in~\cite{Borexino:2003} for the cases of $^{12}$C, $^{13}$C, and $^{16}$O, and in~\cite{DAMA:2000} for the case of $^{136}$Xe.

The focus of this work is a search for the decays of daughter nuclei $^{133}$Sb and $^{133}$Te, which may result from ppp and npp nucleon decay modes, respectively. The subsequent  decays are shown in Fig. \ref{fig:chain}. The Q-Values and half-lives of the decays in Fig. \ref{fig:chain} are given in Table \ref{tab:daughters}. Nucleon decay to $^{133}$I could also produce detectable events, but this decay path is not studied in this work. The final decay, of $^{133}$Xe, is below the analysis energy threshold of 980~keV and is therefore not detectable in this experiment. 

\begin{table}
\caption{Q-values and half-lives of the daughters of triple nucleon decays \cite{Braun:1984}.}
\begin{tabular}{c c c}
\hline
\parbox[t]{1.4cm}{Daughter Isotope}&\parbox[t]{2.8cm}{Q-Value [keV]}&\parbox[t]{1.4cm}{Half-life}\\
\hline
	$^{133}$Sb 	& 4010 & 2.51 min\\
	$^{133}$Te 	& 2955 & 12.5 min\\
	$^{133m}$Te & 3289 & 55.4 min\\
    $^{133}$I	& 1757 & 20.8 h  \\
    $^{133}$Xe	& 427.4& 5.25 d  \\
\hline
\end{tabular}
\label{tab:daughters}
\end{table}

The partial lifetime for all nucleon decay modes to daughter $i$ = ($^{133}$Sb or $^{133}$Te) is:

\begin{equation}\label{eq:tau_Neff}
\tau_i = \frac{N_{nucl} T \epsilon}{S_i},
\end{equation}
where $S_i$ is the number of the observed daughter nuclei of type $i$, $T$ is the experiment livetime, $N_{\rm{nucl}}$ is the number of initial parent nuclei, and $\epsilon$ is the detection efficiency. The lifetime $t_j$ for a particular nucleon decay mode $j$ (e.g. ppp or npp) is given by: 
\begin{equation}
\frac{1}{t_j} = Br_{j} \sum_i \frac{1}{\tau_i}
\end{equation}
where $Br_{j}$ is the total branching ratio of nucleon decay via mode $j$. 

In addition to $^{136}$Xe, EXO-200 contains a non-negligible amount of $^{134}$Xe($\sim$19\% ~\cite{Auger:2012gs}), which can also be utilized to search for nucleon decays using the same strategy. However, given the $\sim$4 times smaller exposure and lower Q-Values of the corresponding daughter isotopes, the resulting lifetime limits are not expected to be competitive with the ones obtained using $^{136}$Xe analysis. Hence this work focuses solely on the $^{136}$Xe analysis.

\section{Detector description}\label{sec:Detector} 
The EXO-200 detector is a cylindrical single-phase time projection chamber (TPC) filled with liquid xenon (LXe) enriched to 80.7\% in \(^{136}\)Xe. A detailed description of the detector is available elsewhere~\cite{Auger:2012gs}. The detector is constructed from components carefully selected to minimize internal radioactivity~\cite{Leonard:2016}. External radioactivity is shielded by 25 cm thick lead walls surrounding the detector on all sides. Additional passive shielding is provided by \(\sim\)50 cm of high-purity cryogenic fluid~\cite{3m} filling the copper cryostat that houses the TPC. The detector is located inside a clean room at the Waste Isolation Pilot Plant (WIPP) in Carlsbad, New Mexico, USA, under an overburden of 1624\(^{+22}_{-21}\) meters water equivalent~\cite{Albert:2014}. The remaining cosmic-rays are detected with 96.0$\pm$0.5\% efficiency by an active muon veto system consisting of plastic scintillation panels surrounding the clean room on four sides. 

Energy deposited in the TPC by ionizing radiation produces free charge and scintillation light, which are registered by anode wire grids and arrays of avalanche photodiodes, respectively. The TPC has a central cathode  at -8~kV and two drift volumes. It allows for three-dimensional position reconstruction of energy depositions. In a given event, charge deposits (``clusters'') that are spatially separated by \(\sim\)1 cm or more can be individually resolved. Events can then be classified as single-site (SS), or multi-site (MS), depending on the number of observed charge clusters. The total energy of an event is determined by combining the charge and scintillation signals. This combination achieves better energy resolution than in each individual channel due to the anticorrelation between them~\cite{Conti:2003}. Radioactive $\gamma$ sources are deployed at several positions near the TPC to characterize the detector response and validate the Monte Carlo simulation.

\section{Experimental data and analysis}\label{sec:Analysis}
A total of $596.70$ live days of data were accumulated for this dataset. The fiducial volume is described by an hexagonal prism with an apothem of 162~mm and length coordinate $1 < |Z| <182 $~mm (with $Z = 0$ corresponding to the cathode location). Nucleon decays occurring in the liquid xenon anywhere in the detector could produce an event in the fiducial volume. Consequently, the $^{136}$Xe mass is 136.5~kg, or 6.05$\times$10$^{26}$ atoms of $^{136}$Xe, resulting in an exposure of 223~kg$\cdot$yr.

Probability density functions (PDFs) for signal and background components are created using a Monte Carlo simulation utilizing Geant4 \cite{GEANT4:2016}. The signal and background PDFs as functions of energy are combined into an overall model and fit to the data.  An additional complication for this analysis arises from the fact that some fraction of the daughters of the decay is ionized and will drift toward the cathode before decaying. Decays occurring on the cathode and in the inactive xenon can be detected via $\gamma$ rays emitted into the fiducial volume. To increase the sensitivity to events occuring on the cathode, the fiducial volume is extended from $|Z|>10$~mm in the 0$\nu$ analysis \citep{Albert:2014b} to $|Z|>1$~mm in this analysis.  To account for daughter drift in the shape of the signal PDF in energy it was necessary to model the spatial distribution of daughter decays in the detector.  This distribution, along with detection efficiencies, determines a probability that a nucleon decay to a
particular daughter in a given detector volume will result in a daughter decay, which produces detected
events in the fiducial volume that pass analysis cuts. These event probabilities for decay to $^{133}$Sb are listed in Table. \ref{tab:distributions}. 

\begin{table}
\caption{Detection probabilities for events originating from decays in different detector volumes.}
\begin{tabular}{llc lc lc lc}
\hline
\parbox[t]{1.75cm}{Daughter} &\parbox[t]{1.75cm}{Active Xe} &\parbox[t]{2cm}{Inactive Xe} &\parbox[t]{1.5cm}{Cathode}\\
\hline
$^{133}$Sb 	& 0.0439(59) & 0.0059 & 0.0116(23) \\
$^{133}$Te 	& 0.0154(42) & 0.0020 & 0.0267(18) \\
$^{133\rm m}$Te & 0.0084(26) & 0.0019 & 0.0161(10)\\
$^{133}$I 	& 0.0035(16) & 0.0004 & 0.0259(6)\\

\hline
\end{tabular}
\label{tab:distributions}
\end{table}

A critical parameter in this distribution model is the ion fraction of the nucleon decay daughter and subsequent $\beta$ decay daughters. Daughter ion fractions for radioactive decays in LXe were measured for the first time recently in EXO-200 \cite{AlphaIon:2015}.  In the $^{222}$Rn decay chain, it was found that the $^{218}$Po daughter ion fraction from $^{222}$Rn $\alpha$ decay is 50.3$\pm$3.0\% and the $^{214}$Bi daughter ion fraction from $^{214}$Pb $\beta$ decay is 76.4$\pm$5.7\%.  As the daughter ion fraction for ppp and npp decay is not known, the above measurements have been used in a simple model as a guideline for an estimate. 

A dominant process leading to the final daughter ion fraction is the interplay of recombination of the daughter ion with the local electron density in the electron cloud from the decay and the rate at which the electron cloud is drawn away from the daughter ion by the electric field. Charge transfer collisions with positively charged holes may also occur in the $\sim$10$^5$ longer time frame during which the cloud of holes is drawn away.  While the details of this process are complex, at the electric field of the detector, the local ionization density at the daughter location should be the critical parameter that determines the final daughter ion fraction.

In $\beta$ decay, there is negligible daughter recoil energy, and the electron cloud is of large radius and low density.  In contrast, in $^{222}$Rn $\alpha$ decay, the $^{218}$Po daughter recoils with 101 keV of energy.  The local electron cloud at the final stopping place of the daughter, due to ionization both from the $\alpha$ particle and the nuclear recoil, is of much smaller radius and much higher density. The greater recombination that follows between daughter ions and the higher electron density explains the smaller observed daughter ion fraction in $\alpha$ decay compared to $\beta$ decay. 

For ppp and npp decay the dominant processes are \cite{Babu:2003}:
\begin{equation}
ppp \rightarrow e^+ + \pi^+ + \pi^+
\end{equation}
\begin{equation}
npp \rightarrow e^+ + \pi^+
\end{equation}
The highly energetic charged particles emitted leave  low ionization density tracks near the daughter location. The recoil energy of the daughter is large, e.g. 15~MeV average for $^{133}$Sb from ppp decay in reaction (3). SRIM simulations of the 3-D ionization density in the neighborhood of the daughter ion indicate a similar shape and density to the electron cloud for $^{136}$Xe ppp and npp decay and $^{222}$Rn $\alpha$ decay \cite{SRIM:2010}.  Thus, a similar daughter ion fraction is expected for ppp and npp decay as for $^{222}$Rn $\alpha$ decays, i.e., $\sim$~50\%.  To confirm this more quantitatively, a simple model of charge drift in the detector field with varying recombination and charge transfer rate parameters, was applied to the initial electron and hole distributions simulated for individual ppp and $^{222}$Rn $\alpha$ decay events.  For a given recombination rate, the charge transfer rate was adjusted to yield a 50.3\% daughter ion fraction on the average in $\alpha$ decay events. With the same pair of parameters, the average daughter ion fraction for ppp events was within 4.3\% of 50\% for a wide range of physically reasonable assumed recombination rates. Conservatively doubling this range to $\pm$9\% for model uncertainty and adding the 3\% experimental uncertainty in quadrature, a daughter ion fraction of 50$\pm$10\% was used in this analysis for the ppp and npp decay. The observed daughter ion fraction of 76$\pm$6\% for $^{214}$Pb $\beta$ decay was used for the subsequent $\beta$ decays. Extreme values of these two daughter ion fractions are used to generate the detection probability uncertainties in Table \ref{tab:distributions}.

The PDF model includes all components used in \cite{Albert:2014b}, except $0\nu\beta\beta$, as well as a component specific to the nucleon decay daughter being investigated. The model is parameterized by the event counts and SS fractions, SS/(MS+SS), of the individual components, as well as by two normalization parameters, which are included in the same way as in \cite{Albert:2014}. A negative log-likelihood function is formed between the data and the overall PDFs for both SS and MS spectra. Five Gaussian constraints are included to incorporate systematic uncertainties determined in stand-alone studies described in the next section. The PDFs for $\beta$-like components also include a $\beta$ scale parameter to account for the potential difference in energy scales of $\beta$-like and $\gamma$-like events~\cite{Albert:2014b}. This parameter is very close to 1 in all fits. For $^{133}$Sb and $^{133}$Te decay chains this provides only an approximate description, as they are comprised of a comparable number of both types of events. This was shown to produce a negligible impact on the final result.

\section{Systematic Errors}
 Systematic uncertainties were accounted for by the same methodology as in \cite{Albert:2014b, EXOMajoron:2014}. They are included in the maximum likelihood fit as nuisance parameters, constrained by normal distributions. The radon activity in the liquid xenon and relative fractions of neutron-capture related PDF components were constrained by 10\% and 20\%, respectively, as evaluated in \citep{Albert:2014}. The SS fraction uncertainty was determined to be 4\% in a similar way as in \cite{Albert:2014}.

The error on signal detection efficiency was re-evaluated from that used in \cite{EXOMajoron:2014} to account for the change in the fiducial volume. Studies of the rate agreement using calibration sources indicated an error increase of 1\%. This was added in quadrature to the previously determined value resulting in a total error of 9\%.

Three sources of uncertainty contributed to the error associated with shape differences between data and Monte Carlo: discrepancies in the energy spectra between data and Monte Carlo, choice of components in the background model, and ion fraction uncertainties. To estimate the first, a linear function of energy was fit to the difference between data and Monte Carlo for calibration sources in the range 1000-2000~keV and was used to unskew the background PDFs: $^{226}$Ra source fit for U and Rn backgrounds, $^{60}$Co source fit for Co backgrounds, and $^{228}$Th source fit for all other $\gamma$-like backgrounds. Toy Monte Carlo datasets generated from this model with a variable number of signal events injected were fit with the normal background and signal PDFs. For decay to $^{133}$Sb, the fractional difference between fit and injected
number of signal events was roughly constant at 7\%. For $^{133}$Te the difference was linear at small number
of injected events ($<$5000) and approached a constant difference of around 5000 at high number of injected
events.


To evaluate the uncertainty associated with the background model, decays of Th, U, and Co were simulated in different locations than that in the default model. Changes in the result due to the substitutions were added into the shape differences in quadrature along with the ion fraction uncertainties to give a total value for this error of 11\%. The robustness of the fits was also checked, in a manner analogous to \cite{EXOMajoron:2014}, against the existence of $^{88}$Y and metastable $^{110}$Ag in the vessel material. There was no significant impact on the results.

\section{Results and conclusion}\label{sec:Results} 

\begin{figure*}[htbp]
\centering
\includegraphics[scale = .45]{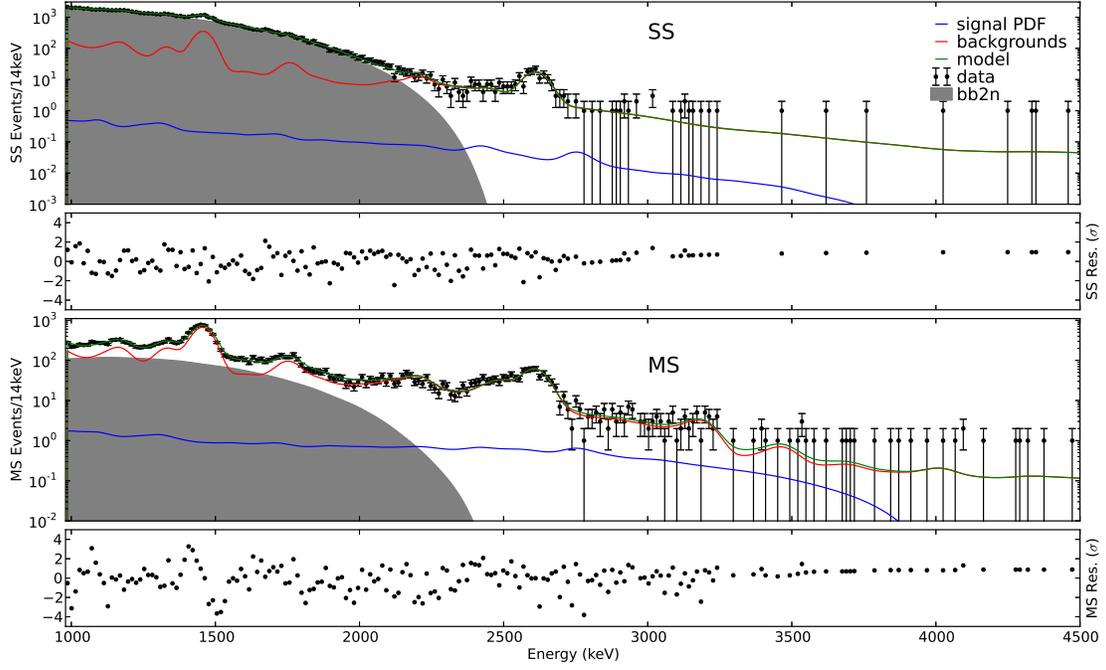}
\caption{SS and MS data, the best-fit model for decay to $^{133}$Sb and residuals. The dominant component, $2\nu\beta\beta$ decay, is shown in gray.}	
\label{fig:fit}
\end{figure*}

\begin{figure*}[htbp]
\centering
\includegraphics[scale = .45]{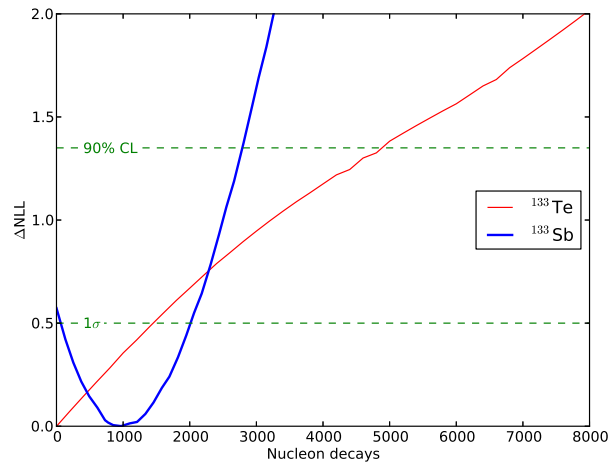}
\caption{Likelihood profiles for decays to $^{133}$Sb and $^{133}$Te. The dashed lines represent the 1$\sigma$ and 90\% confidence limits (C.L.).}	
\label{fig:profiles}
\end{figure*}

The experimental SS and MS energy spectra and the best fit, performed simultaneously for SS and MS events, is shown in Fig.~\ref{fig:fit} for nucleon decay to $^{133}$Sb. The results of profile-likelihood scans, performed for each daughter decay, are shown in Fig.~\ref{fig:profiles}. No statistically significant signal is observed in either case. For decay to $^{133}$Te the best fit signal is zero. The 90\% CL upper limits were derived under the assumption of Wilks' theorem \cite{Wilks:1938,Cowan:1998}, given the large statistics of the data set.  They are 2.8$\times$10$^3$ events for $^{133}$Sb and 4.9$\times$10$^3$ events for decay to $^{133}$Te, corresponding to a lifetime limit of 3.3$\times$10$^{23}$ yrs for nucleon decay to $^{133}$Sb and a lifetime limit of 1.9$\times$10$^{23}$ yrs for nucleon decay to $^{133}$Te. The efficiency factor of 0.93 used to calculate the lifetimes in Eq. \ref{eq:tau_Neff} accounts for a cut of data events occurring within 1~s of a previous event. This cut is used to discriminate against correlated events in the detector. All other efficiencies are incorporated via the signal weights in Table. \ref{tab:distributions} and do not need to be applied after the fit.

In conclusion, we report results from a search for baryon number violating decays in $^{136}$Xe using two years of data from EXO-200. The results do not show statistically significant evidence for triple-nucleon decay to either $^{133}$Sb or $^{133}$Te. The lifetime limits obtained on the decays are the most stringent to-date surpassing the previous results~\cite{DAMA:2006} by a factor of 9 and 7 for $^{133}$Sb and $^{133}$Te respectively.

\section*{Acknowledgments}
    EXO-200 is supported by DOE and NSF in the United States, NSERC in Canada, SNF in Switzerland, Institute fo Basic Science (IBS) in Korea, Russian Foundation for Basic Research (RFBR) in Russia, DFG in Germany, and CAS and International Science and Technology Cooperation Project (ISTCP) in China. EXO-200 data analysis and simulation uses resources of the National Energy Research Scientific Computing Center (NERSC). We gratefully acknowledge the KARMEN collaboration for supplying the cosmic-ray veto detectors, and the WIPP for their hospitality.

\bibliography{ND5_2}

\end{document}